\documentclass[3p,times,twocolumn]{elsarticle}

\usepackage{ecrc}
\usepackage{atlasphysics}
\usepackage{amsmath}
\usepackage{graphicx}
\usepackage{url}
\usepackage{gensymb}
\usepackage{amssymb}

\newcommand{\powheg}{\texttt{POWHEG}}
\newcommand{\jetphox}{\texttt{jetphox}}
\newcommand{\taa}{\langle T_{\mathrm{AA}} \rangle}
\newcommand{\mNpart}{\langle N_{\mathrm{part}} \rangle}
\newcommand{\mNcoll}{\langle N_{\mathrm{coll}} \rangle}

\volume{00}

\firstpage{1}

\journalname{Nuclear Physics B}

\runauth{T. Balestri}

\jid{nppp}

\jnltitlelogo{Nuclear Physics B}

\usepackage{amssymb}

\biboptions{square,comma,numbers,sort&compress}
\usepackage[figuresright]{rotating}

\begin{document}

\begin{frontmatter}



\dochead{}

\title{Electroweak bosons in Pb+Pb and $p$+Pb collisions}

\author[label1]{Thomas Balestri on behalf of the ATLAS Collaboration}
\address[label2]{Stony Brook University, NY 11794, USA}

\begin{abstract}
    Electroweak boson ($\Wboson$, $\Zboson$, $\gamma$) measurements in Pb+Pb collisions at $\sqrt{s_{NN}}=2.76\TeV$ and in $p$+Pb collisions at $\sqrt{s_{NN}}=5.02\TeV$ are presented with the ATLAS detector at the LHC. In Pb+Pb, electroweak boson yields are shown to be independent of centrality. Differential measurements in absolute pseudorapidity are used to investigate nuclear effects to the free-proton parton distribution function (PDF). The distributions lack the experimental precision to unambiguously identify the presence of nuclear modifications. In $p$+Pb,
    the $\Zboson$ boson cross section is measured as a function of center-of-mass rapidity $y_{Z}^{*}$ and the momentum fraction of the lead-going parton (Bjorken $x_{Pb}$). The distributions are asymmetric and model predictions underestimate the data at large $x_{Pb}$. The overall shape is best described by including nuclear effects. The differential cross section is also measured in different centrality classes and shows evidence of spatially-dependent nuclear PDFs. The $\Zboson$ boson production yields are measured as a function of the mean number of
    participants using Glauber and Glauber-Gribov color fluctuation models. Binary scaling of the yields is observed utilizing the standard Glauber model after applying a centrality-bias correction.  
\end{abstract}


\end{frontmatter}

\section{Introduction}
\label{sec:1}

Two major questions in high-energy nuclear physics are: how are free nucleon parton distribution functions (PDFs) modified in heavy-nuclear systems and do we understand the collision geometry and centrality in Pb+Pb and $p$+Pb collisions. These questions can be addressed by studying the production yields and kinematic properties of electroweak bosons (i.e. $\gamma$, $\Wboson$, $\Zboson$). Electroweak (EW) bosons do not interact via the strong force and thus their yields are not
expected to be affected by the presence of a quark-gluon plasma (QGP). This characteristic offers a means to test various models of the collision geometry. Furthermore, EW bosons are sensitive to modifications of the free nucleon PDF and thus may be used to detect the presence and extent of nuclear effects. The type of modification is two-fold: the trivial modification due to the different valence quark content of the proton and neutron (so-called isospin effects) and Bjorken-$x$ dependent
modifications that include shadowing, antishadowing, EMC-effect, and Fermi-motion. The EPS09 nuclear PDF set~\cite{Eskola:2009uj} incorporates these nuclear effects into their PDF calculations and also considers the possibility of spatially-dependent nuclear modifications.           
The following sections present EW boson analyses conducted in Pb+Pb collisions at $\sqrt{s_{NN}}=2.76\TeV$ and in $p$+Pb collisions at $\sqrt{s_{NN}}=5.02\TeV$ with the ATLAS detector~\cite{Aad:2008zzm} at the LHC.

\section{Photons}
\label{sec:2}

\begin{figure}[!htbp]
\begin{center}
  \resizebox{0.5\textwidth}{!}{
    \includegraphics[]{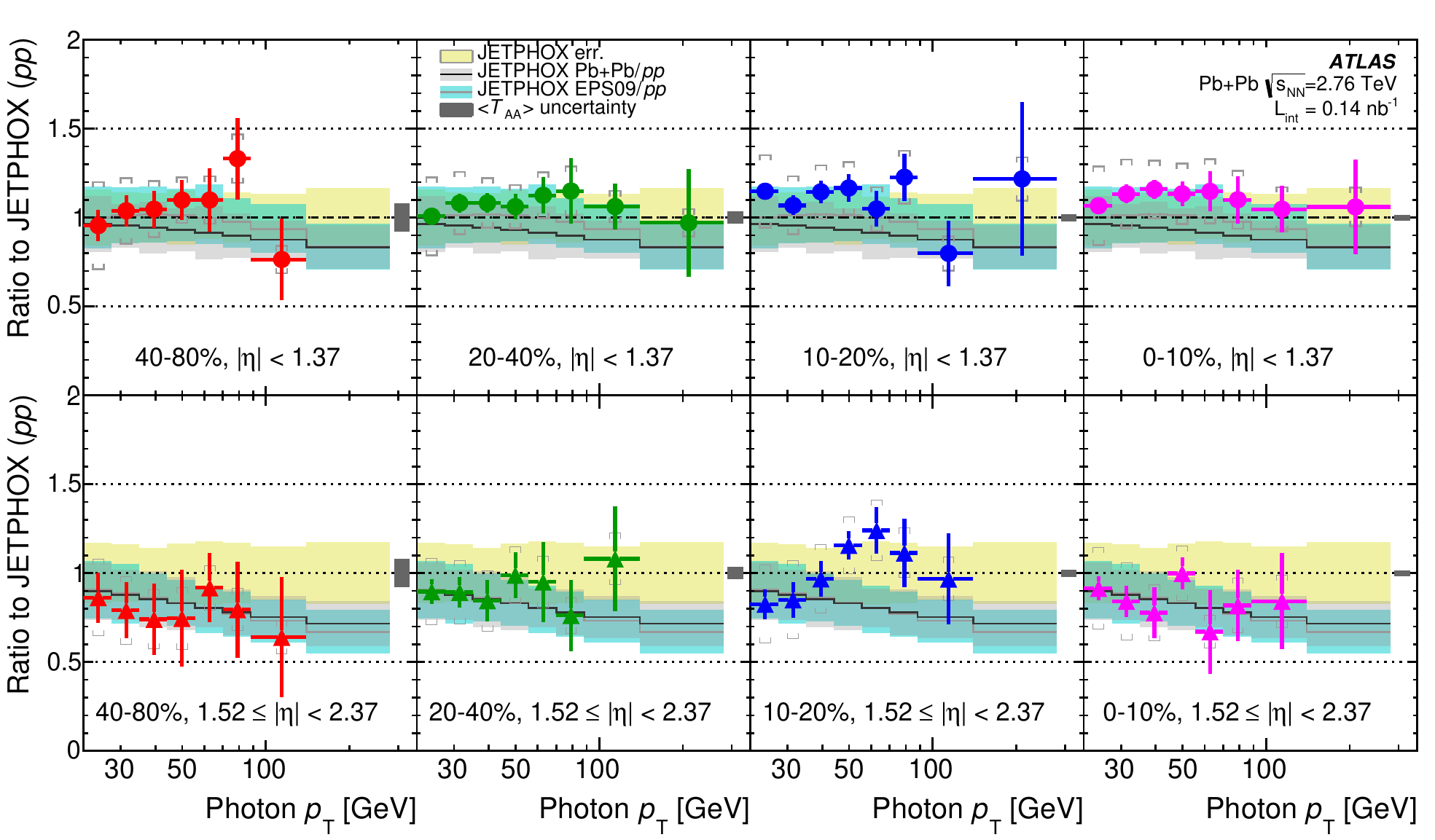}
  }
\caption{\label{fig:1}
    Ratio of fully corrected prompt photon yields in Pb+Pb and photon yields from NLO pQCD calculations in $pp$ at $\sqrt{s_{NN}}=2.76\TeV$ as a function of photon $\pt$. The top panel shows ratios in $|\eta|<1.37$ and the bottom panel in $1.52\leq|\eta|<2.37$. Predictions from \jetphox in Pb+Pb collisions with and without EPS09 nuclear corrections are also shown in each figure. Statistical uncertainties are shown by the bars and systematic uncertainties by upper and lower braces. Scaling
   uncertainties from the average nuclear thickness function $\taa$ are also shown~\cite{Aad:2015lcb}.     
}
\end{center}
\end{figure}

\begin{figure}[!htbp]
\begin{center}
  \resizebox{0.5\textwidth}{!}{
    \includegraphics[]{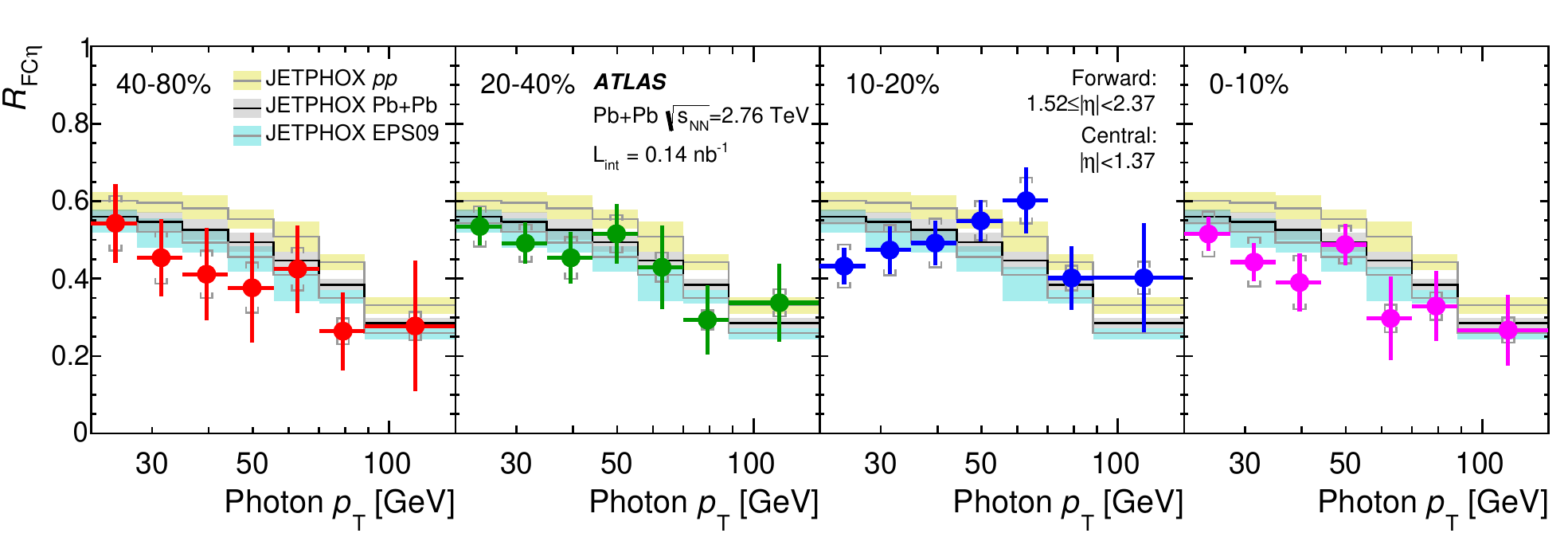}
  }
\caption{\label{fig:2}
   Ratio of fully corrected prompt photon yields in Pb+Pb at $1.52\leq|\eta|<2.37$ and photon yields at $|\eta|<1.37$ at $\sqrt{s_{NN}}=2.76\TeV$ as a function of photon $\pt$. 
   Predictions from \jetphox in Pb+Pb collisions with and without EPS09 nuclear corrections are also shown in each figure. Statistical uncertainties are shown by the bars and systematic uncertainties by upper and lower braces~\cite{Aad:2015lcb}. 
}
\end{center}
\end{figure}

Figures~\ref{fig:1} and ~\ref{fig:2} show measurements used to study nuclear modifications of the PDF using photons in Pb+Pb collisions. These results are published in Ref.~\cite{Aad:2015lcb}. Figure~\ref{fig:1} presents the ratios of the photon yields from the data in Pb+Pb collisions and from a next-to-leading-order (NLO) pQCD prediction in $pp$ collisions as a function of photon $\pt$. These ratios are shown at central and forward rapidity intervals and are segmented into four
different centrality classes. Also shown in Figure~\ref{fig:1} are the ratios of
the photon yields in simulated Pb+Pb collisions with and without EPS09 nuclear effects. Within the uncertainty of the measurement, the models without nuclear effects cannot be excluded. 

Figure~\ref{fig:2} presents the forward-to-central ratio of the yields in the data and in simulated Pb+Pb collisions with and without EPS09 nuclear effects as a function of photon $\pt$. These are also shown for four different centrality classes. The forward-to-central yield ratio is more sensitive to nuclear modifications since several systematic uncertainties are correlated in pseudorapidity and thus cancel in the ratio. However, the measurement lacks the precision to
indicate the presence of nuclear modifications.   

\section{$\Wboson$ Boson}
\label{sec:3}
\begin{figure}[!htbp]
\begin{center}
  \resizebox{0.34\textwidth}{!}{
    \includegraphics[]{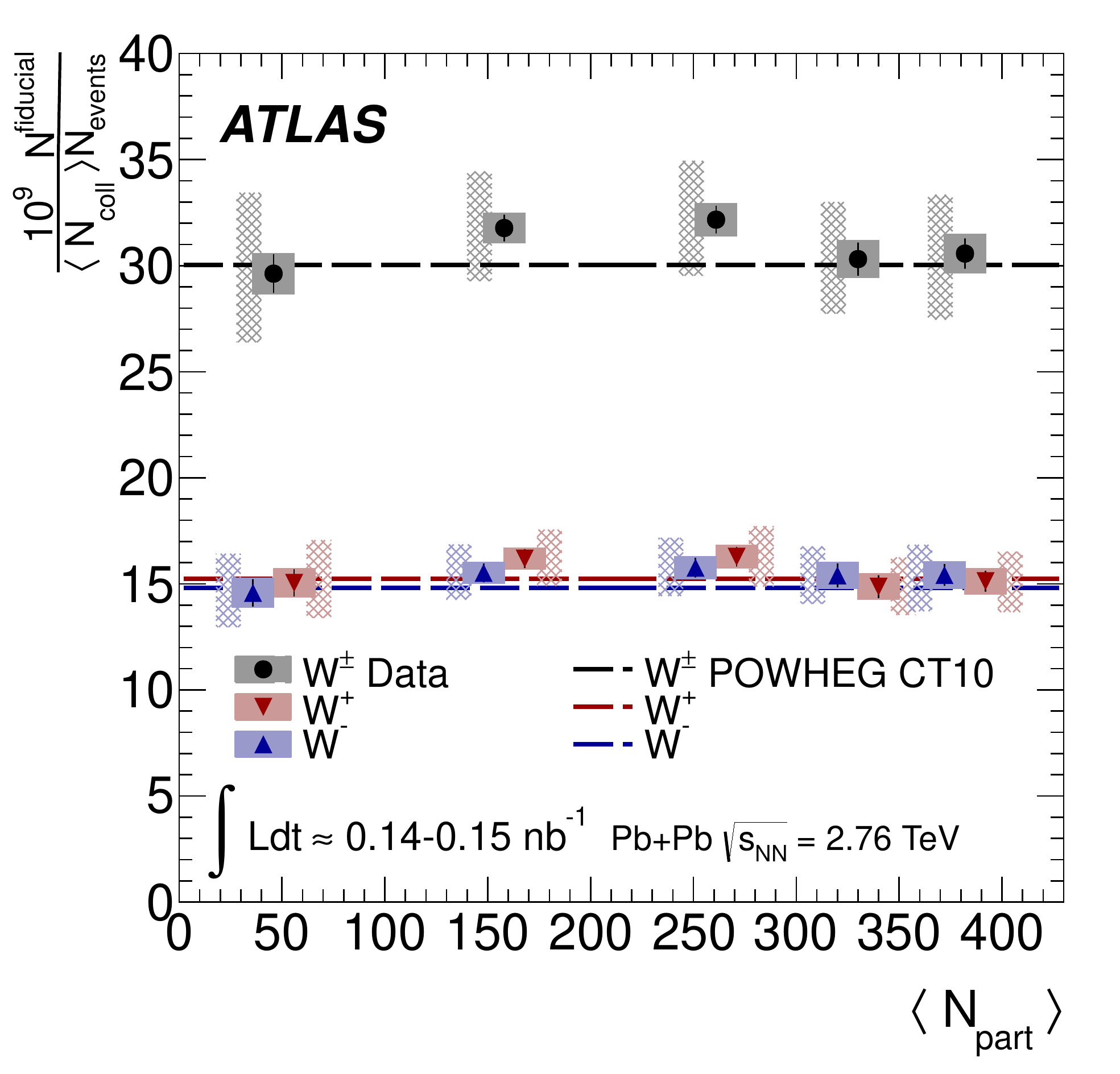}
  }
\caption{\label{fig:3}
    $\Wboson$ boson production yields per binary collision as a function of $\mNpart$. The yields are shown for $\Wboson^{+}$, $\Wboson^{-}$, and $\Wboson^{\pm}$ for the combined electron and muon decay channels. The yields are shown within a fiducial region defined in Ref.~\cite{Aad:2014bha}. Bin-correlated systematic errors are shown as hatches whereas bin-uncorrelated errors are shown as the filled error bars. Solid bars represent statistical uncertainties. Also shown are NLO QCD
    predictions generated with $\powheg$.  
}
\end{center}
\end{figure}

Figure~\ref{fig:3} presents the $\Wboson$ boson production yields per binary nucleon-nucleon collision ($\mNcoll$) as a function of the mean number of participants ($\mNpart$) in Pb+Pb collisions. The yields are obtained from both the electron and muon decay channels and are shown for each leptonic charge class: $\ell^{+}$, $\ell^{-}$, and $\ell^{\pm}$. The $\Wboson$ boson production yields are independent of the size of the overlap collision system and thus scale with $\mNcoll$. Given that
$\mNcoll$ is obtained in each
centrality class using the standard Glauber formulism, this result suggests that the collision geometry can be sufficiently described using the standard Glauber model. As will be shown in the next section, this may not be the case in a $p$+Pb system.    

\begin{figure}[!htbp]
\begin{center}
  \resizebox{0.34\textwidth}{!}{
    \includegraphics[]{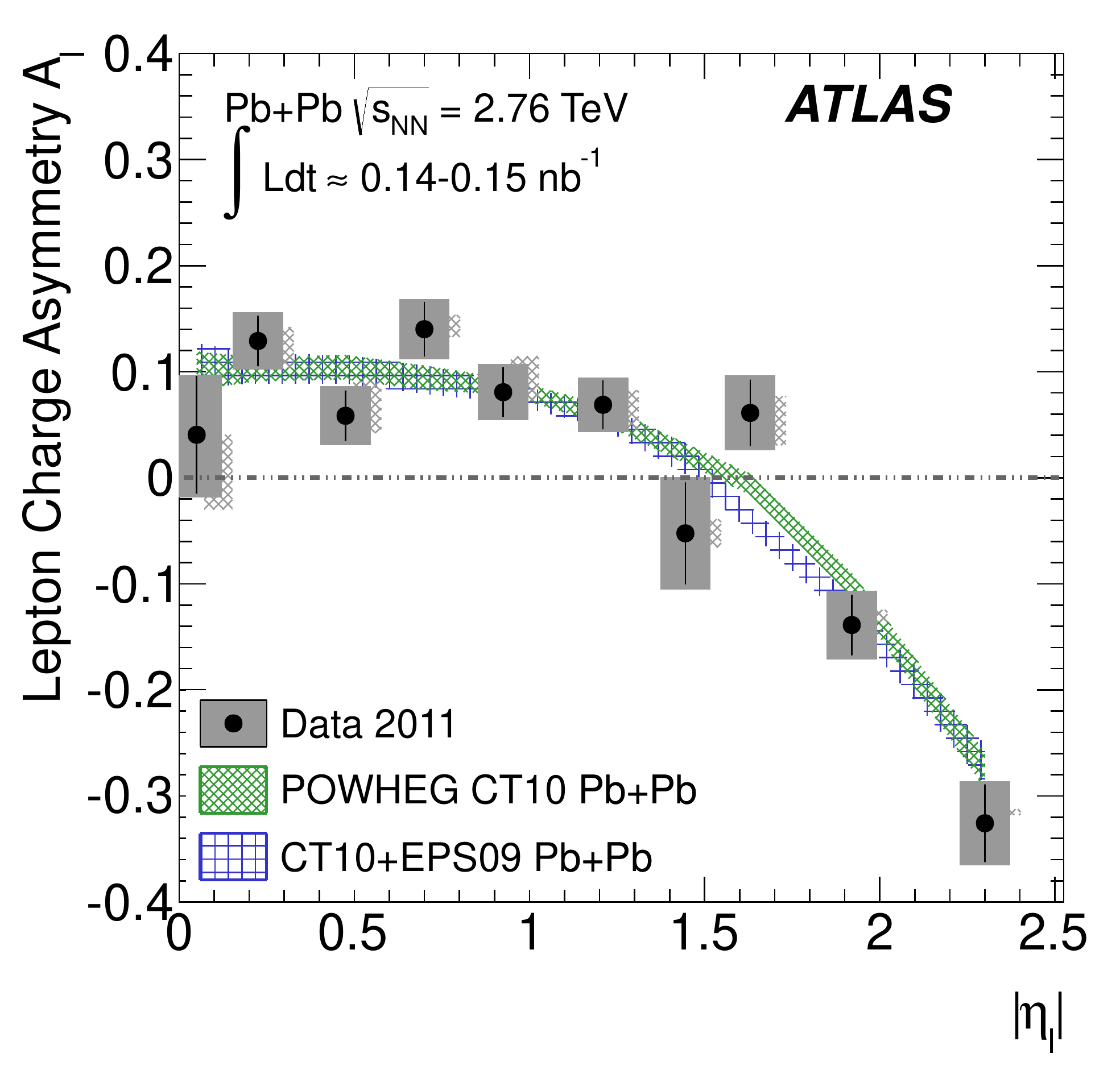}
  }
\caption{\label{fig:4}
    Lepton charge asymmetry as a function of absolute pseudorapidity of the charged lepton. The measurement includes results from both the muon and electron decay channels and the yields are defined within a fiducial region defined in Ref.~\cite{Aad:2014bha}. Bin-correlated errors are shown as hatches, statistical uncertainties as solid bars, and bin-uncorrelated uncertainties added in quadrature to the statistical uncertainties as filled errors bars. Predictions from the CT10 and
    CT10+EPS09 PDFs are also shown for comparison.
}
\end{center}
\end{figure}

Figure~\ref{fig:4} shows the lepton charge asymmetry as a function of lepton absolute pseudorapidity. This observable is sensitive to PDFs and so may also be sensitive to nuclear modifications. To test whether this is the case, predictions from CT10 with and without EPS09 nuclear corrections are provided. The data are consistent with both predictions. Therefore, without further constraining the models and reducing the experimental uncertainties, this measurement is
unable to provide clear evidence
of nuclear modifications to the PDF. A more detailed discussion of these results may be found in Ref.~\cite{Aad:2014bha}. 

\section{$\Zboson$ Boson}
\label{sec:4}
Measurements of $\Zboson$ boson production in Pb+Pb collisions~\cite{Aad:2012ew} validate the conclusions made from the $\Wboson$ boson results and will not be discussed further. Rather, this section will address $\Zboson$ boson production in $p$+Pb collisions~\cite{Aad:2015gta} at $\sqrt{s_{NN}}=5.02\TeV$. 

\begin{figure}[htbp]
\begin{center}
  \resizebox{0.34\textwidth}{!}{
    \includegraphics[]{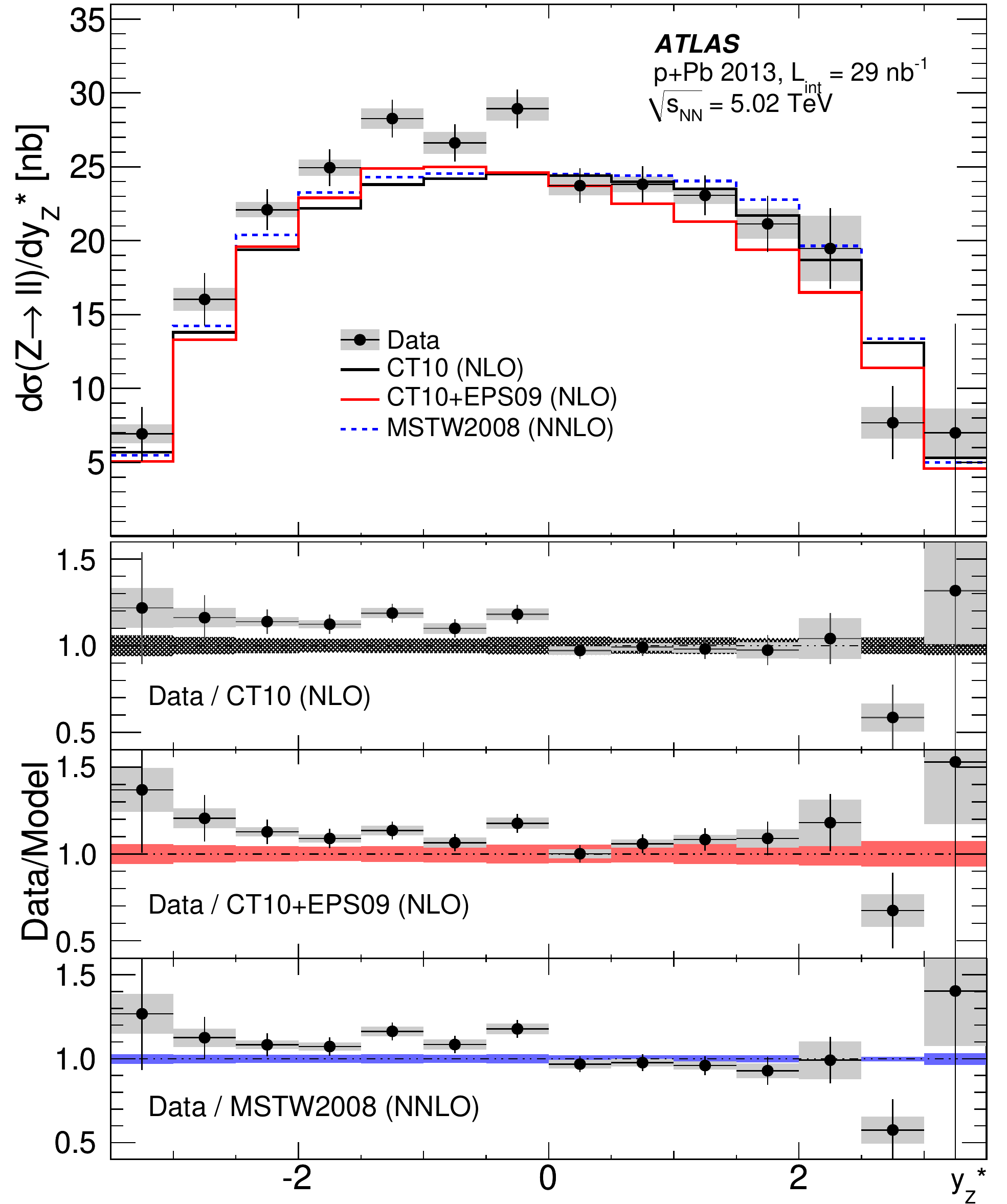}
  }
\caption{\label{fig:5}
    $d\sigma/d y^{*}_{Z}$ distribution in $\Zboson\rightarrow\ell\ell$ events. Several model calculations are also included. The lower panels show the ratios of data to model. Statistical errors are shown as bars and systematic errors as shaded boxes~\cite{Aad:2015gta}.  
}
\end{center}
\end{figure}

\begin{figure}[htbp]
\begin{center}
  \resizebox{0.34\textwidth}{!}{
    \includegraphics[]{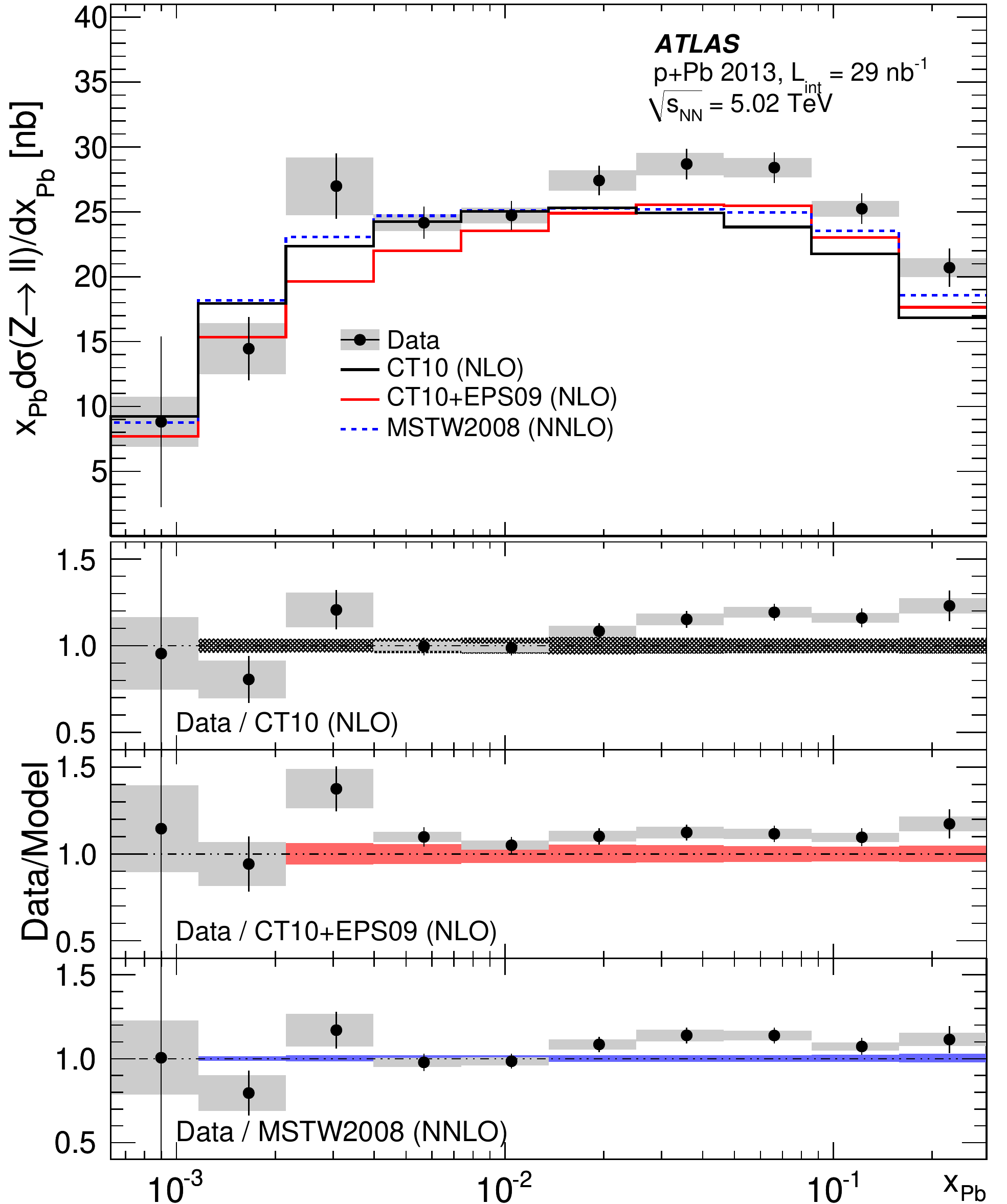}
  }
\caption{\label{fig:6}
    $d\sigma/d x_{Pb}$ distribution multiplied by Bjorken $x$ of the parton in the lead nucles in $\Zboson\rightarrow\ell\ell$ events as a function of $x_{Pb}$. Several model calculations are also included. The lower panels show the ratios of data to model. Statistical errors are shown as bars and systematic errors as shaded boxes~\cite{Aad:2015gta}.
}
\end{center}
\end{figure}

Figure~\ref{fig:5} presents the differential cross-section as a function of the center-of-mass rapidity of the $\Zboson$ boson ($y_{\Zboson}^{*}$). Models using the CT10 PDF set with and without EPS09 nuclear corrections and MSTW2008 are also shown. The data is asymmetric about the center-of-mass, and the models underestimate the data at backward rapidity. This implies that there is an enhancement at large $x_{Pb}=M_{\Zboson}e^{-y_{\Zboson}^{*}}/\sqrt{s_{NN}}$ (see
Figure~\ref{fig:6}). Ignoring the scale, however, incorporation of nuclear effects (i.e.
EPS09) best describes the shape of the data.  

\begin{figure}[!htbp]
\begin{center}
  \resizebox{0.3\textwidth}{!}{
    \includegraphics[]{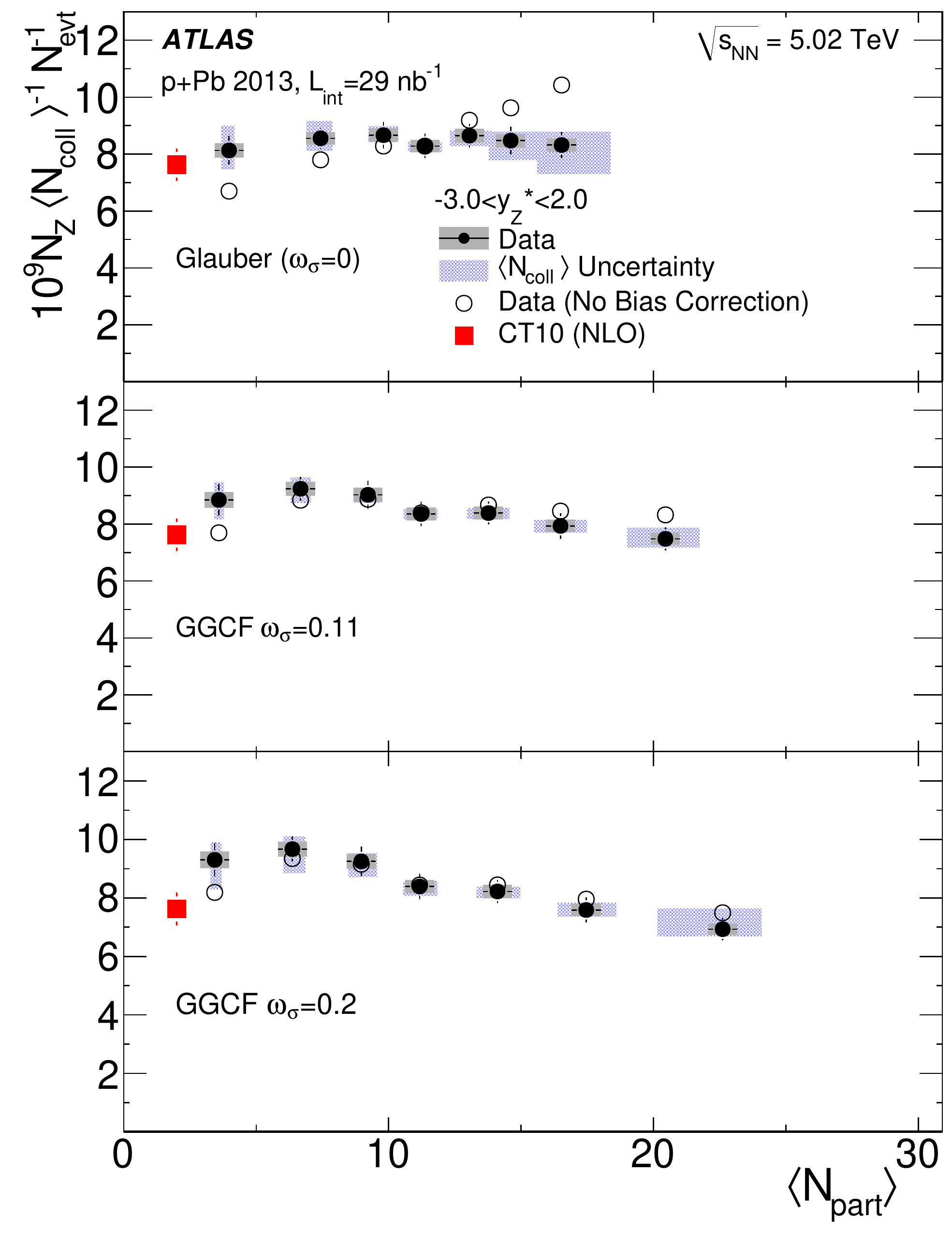}
  }
\caption{\label{fig:7}
    The yield of Z bosons per event scaled by the mean number of nucleon-nucleon collisions $\mNcoll$ as a function of the mean number of participating nucleons $\mNpart$. Each panel uses a different Glauber model configuration in calculating $\mNcoll$ and $\mNpart$. The models used from top to bottom are: standard Glauber, GGCF with $\omega_{\sigma}=0.11$, and GGCF with $\omega_{\sigma}=0.2$. The bars indicate statistical uncertainty and the shaded boxes systematic
    uncertainty~\cite{Aad:2015gta}.
}
\end{center}
\end{figure}

Figure~\ref{fig:7} presents $\Zboson$ boson yields per binary nucleon-nucleon collision as a function of $\mNpart$ for three different geometric models. The Glauber-Gribov Color Fluctuation (GGCF) model considers event-by-event fluctuations in the nucleon-nucleon cross section, and the magnitude of these fluctuations is represented by $\omega_{\sigma}$. Also taken into account in this measurement is a centrality bias introduced by a correlation between a hard-scattering process and underlying-event
activity~\cite{Perepelitsa:2014yta}. Under the assumption that EW boson yields are centrality independent, the results show that with the centrality-bias correction, the standard Glauber model is a better description of the collision geometry. However, analagous measurements using inclusive charged particles in minimum bias collisions~\cite{Aad:2015zza} have shown similar behavior to that seen in the $\Zboson$ boson measurements without applying a centrality bias correction. Thus, which model is a more valid
description of the collision geometry is still under debate.      

\begin{figure}[!htbp]
\begin{center}
  \resizebox{0.3\textwidth}{!}{
    \includegraphics[]{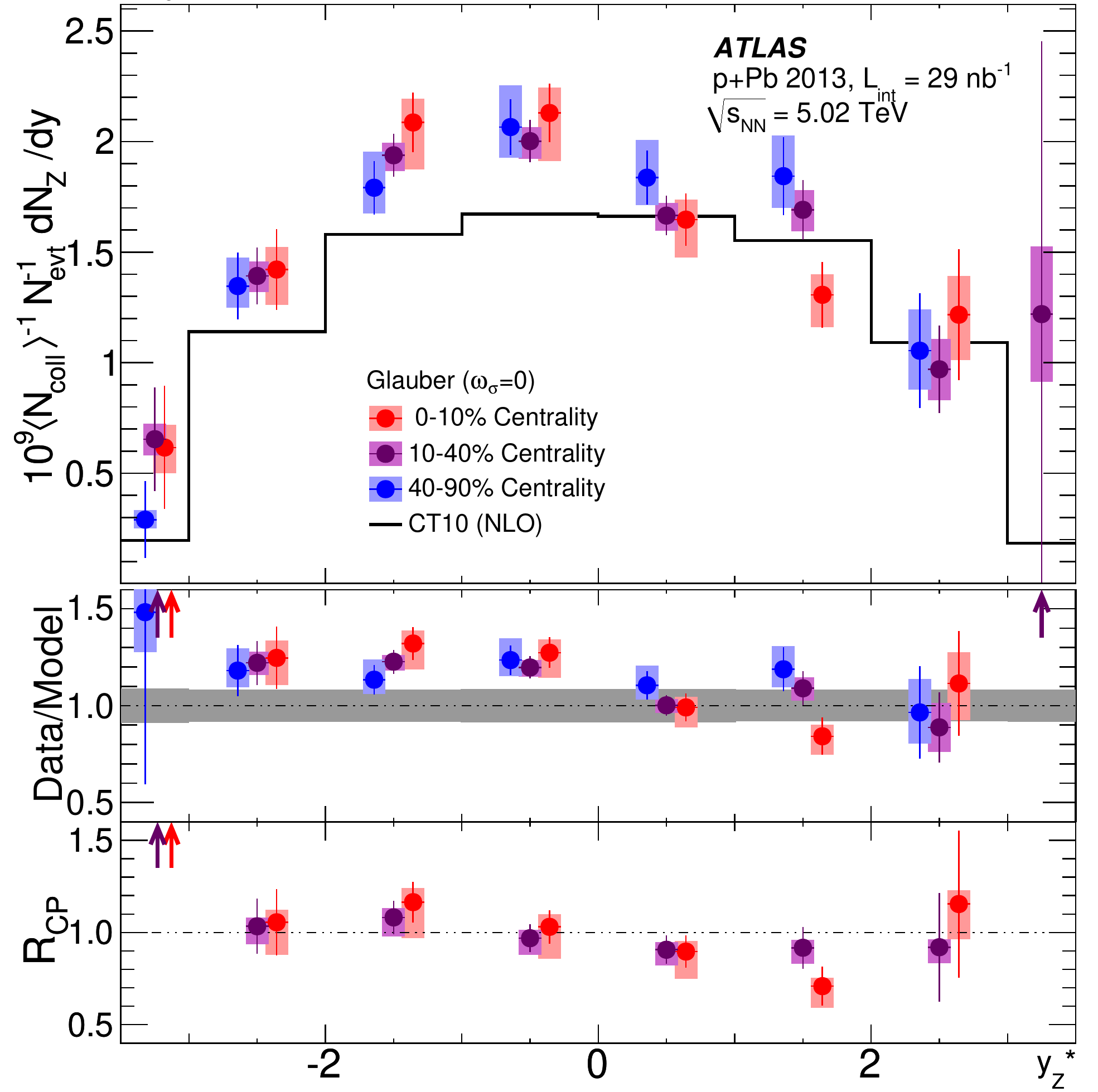}
  }
\caption{\label{fig:8}
    Top panel: The rapidity differential $\Zboson$ boson yields, scaled by $\mNcoll$ for three centrality classes. The data are compared to a CT10 model calculation without nuclear effects. The bars indicate statistical uncertainty and the shaded boxes systematic uncertainty. Bottom panel: $R_{CP}$ for $0-10\%$ and $10-40\%$ centrality classes with respect to the $40-90\%$ centrality class. Arrows indicate values outside the plotted axes~\cite{Aad:2015gta}.
}
\end{center}
\end{figure}

As previously mentioned, EPS09 predicts a possible spatial dependence of the nuclear modifications. To investigate this empirically, Figure~\ref{fig:8} presents the differential cross sections in three centrality classes. The bottom panel shows the yields in each centrality class with respect to those in the most peripheral class ($R_{CP}$). The $R_{CP}$ in the most central class shows a slight rapidity dependence with a slope of $-0.11\pm0.04$, whereas the $R_{CP}$ in mid-central
collisions is independent of $\Zboson$ boson rapidity. This result may be the first experimental hints of spatially-dependent nuclear effects.  

\section{Summary}
\label{sec:5}
Measurements of EW bosons were presented in Pb+Pb collisions at $\sqrt{s_{NN}}=2.76\TeV$ for photons and $\Wboson$ bosons and in $p$+Pb collisions at $\sqrt{s_{NN}}=5.02\TeV$ for $\Zboson$ bosons. The results in Pb+Pb show that the collision geometry is understood with the standard Glauber formulation. However, better experimental precision and further constraints on the PDF sets are needed to support evidence of nuclear effects to the free-nucleon PDF. The cross section for $\Zboson$ bosons was
also presented. The integrated cross section shows an
enhancement at large $x_{Pb}$ relative to model predictions. The cross sections in different centrality classes show signs of spatial dependence of the nuclear PDF. The centrality-dependent $\Zboson$ boson production yields were used to differentiate between the standard Glauber and GGCF models. Binary scaling in the yields was observed using standard Glauber, however which model is most valid is still under consideration. These results have shown the that EW bosons are valuable tools to study
high-energy nuclear collisions. As more data are collected at higher energies, EW boson measurements may serve an increasing role in understanding the physics of heavy-nuclear systems. This research is supported by NSF under grant number PHY-1305037.        




\section*{\refname}
\bibliographystyle{elsarticle-num}
\bibliography{Balestri_T}







\end{document}